\documentclass[10pt,aas2pp4,tighten]{aastex}
\usepackage{color}
\begin{document}

\newcommand{\Wb}{W$_0^{\lambda2796}$}
\newcommand{\Wr}{W$_0^{\lambda2803}$}
\newcommand{\Wl}{W$_0^{lim}$}

\title{The Pittsburgh Sloan Digital Sky Survey \\ \ion{Mg}{2} Quasar Absorption-Line Survey Catalog}

\author{Anna M. Quider \altaffilmark{1}}
\email{aquider@ast.cam.ac.uk}
\and
\author{Daniel B. Nestor \altaffilmark{2}, David A. Turnshek \altaffilmark{3}, \\ Sandhya M. Rao \altaffilmark{3}, 
Eric M. Monier \altaffilmark{4}, Anja N. Weyant\altaffilmark{3}, and Joseph R. Busche\altaffilmark{5}}
\affil{\altaffilmark{1}Institute of Astronomy, Madingley Rd, Cambridge CB3 0HA, UK}
\affil{\altaffilmark{2}Department of Physics \& Astronomy, University of California, Los Angeles, CA 90095}
\affil{\altaffilmark{3}Department of Physics \& Astronomy, University of Pittsburgh, Pittsburgh, PA 15260}
\affil{\altaffilmark{4}Department of Physics, SUNY College at Brockport, NY 14420}
\affil{\altaffilmark{5}Department of Physics, Wheeling Jesuit University, Wheeling, WV 26003}

\begin{abstract}
We present a catalog of intervening \ion{Mg}{2} quasar absorption-line
systems in the redshift interval $0.36 \le z \le 2.28$. The catalog was
built from Sloan Digital Sky Survey Data Release Four (SDSS DR4)
quasar spectra.  Currently, the catalog contains $> 17,000$
measured \ion{Mg}{2} doublets.  We also present data on the $\sim
44,600$ quasar spectra which were searched to construct the catalog,
including redshift and magnitude information, continuum-normalized
spectra, and corresponding arrays of redshift-dependent minimum rest
equivalent widths detectable at our confidence threshold. The catalog is
available on the web. A careful second search of 500 random spectra indicated that, 
for every 100 spectra searched, approximately one significant \ion{Mg}{2}
system was accidentally rejected. 
Current plans to expand the catalog beyond DR4 quasars are discussed. 
Many \ion{Mg}{2} absorbers are known to be
associated with galaxies. Therefore, the combination of large size and
well understood statistics makes this catalog ideal for precision
studies of the low-ionization and neutral gas regions associated with
galaxies at low to moderate redshift.  An analysis of the statistics
of \ion{Mg}{2} absorbers using this catalog will be presented in a
subsequent paper.
\end{abstract}

\keywords{galaxies: ISM --- quasars: absorption lines}

\section{Introduction}

Quasar absorption line (QAL) surveys form the basis for selecting and
studying large numbers of cosmologically-distant galaxies via their
gas cross sections.  Of particular usefulness among QALs is the
\ion{Mg}{2}$\lambda\lambda$2796,2803 resonance doublet.
{\mbox{Mg$^+$}} is a dominant ionization stage in gas having a large
neutral hydrogen fraction, and the
\ion{Mg}{2}$\lambda\lambda$2796,2803 transitions have particularly
strong oscillator strengths, making the doublet an excellent tracer of
low-ionization and neutral gas.   Additionally, in comparison to other
UV absorption lines commonly found in quasar spectra, the relatively
long rest wavelength of the doublet permits its identification with
ground-based optical spectroscopy down to fairly low redshift ($z
\gtrsim 0.14$; e.g., Nestor, Turnshek, \& Rao 2006).

The utility of \ion{Mg}{2}\ absorbers for the study of galaxies has
prompted several QAL surveys in past decades. Initial studies were
usually made at moderate spectral resolution and signal-to-noise
ratio, and this generally resulted in the detection of absorption
systems with moderate to strong rest equivalent widths,  \Wb $\gtrsim
0.3$ \AA. The first surveys (e.g., Weymann et al.\ 1979; Lanzetta, Turnshek,
\& Wolfe 1987; Tytler et al.\ 1987; Sargent, Steidel, \&
Boksenberg 1988; Caulet 1989)  contained a few to a few dozen systems.
The survey of Steidel \& Sargent (1992), which identified more than
100 \ion{Mg}{2}\ absorption systems, was the largest statistical
sample of \ion{Mg}{2} systems for more than a decade.  Imaging surveys
along quasar sightlines demonstrate that the bulk of the \ion{Mg}{2}
absorbers are associated with galaxies (e.g., Bergeron \& Boiss\'{e}
1991; Steidel, Dickinson, \& Persson 1994; Kacprzak et al. 2010; Chen et al. 2010; 
Rao et al. 2011).  Thus, the \ion{Mg}{2} absorption-line
catalogs have been used by various authors to study subsamples of
systems with the aim of investigating the properties of the
low-ionization and neutral gas regions associated with galaxies. These
\ion{Mg}{2}-based studies have the potential to reveal information on
galactic halos/disks (e.g., Charlton \& Churchill 1996), outflows
induced by star formation and supernovae (e.g., Norman et al. 1996;
Bond et al. 2001; Nestor et al. 2010), accreting protogalactic gas (Mo
\& Miralda-Escude 1996), merger-induced activity and tidal debris
(e.g., Sargent \& Steidel 1990; Maller et al. 2001; Kacprzak et
al. 2007), dwarf galaxies (e.g., LeBrun et al. 1993), low surface
brightness galaxies (e.g., Phillips, Disney, \& Davis 1993), and even
the extended environments of other quasars (e.g., Bowen et
al. 2006). High-resolution spectroscopy is generally required to
resolve the kinematic ``component'' structure of the absorbing gas;
this component structure provides important detailed information on
the velocity field of the gas associated with an absorbing galaxy
(e.g., Churchill et al. 2000; Churchill \& Vogt 2001; Steidel et
al. 2002).   It turns out that a significant fraction of strong \ion{Mg}{2}
systems (\Wb $ \ge 0.5$ \AA) are damped Ly$\alpha$ (DLA) absorbers
with neutral gas column densities $N_{HI} \ge 2\times10^{20}$ atoms
cm$^{-2}$, and weaker \ion{Mg}{2} systems are not DLAs (e.g., Rao,
Turnshek, \& Briggs 1995; Rao \& Turnshek 2000; Rao, Turnshek, \&
Nestor 2006). Thus, \ion{Mg}{2}-based surveys are particularly
effective and important for preselecting systems to be used in DLA
surveys at redshifts $z \lesssim 1.65$, when the Ly$\alpha$ line falls
in the UV and is unobservable from the ground.  For example, follow-up
imaging of DLA galaxies shows that DLAs trace the neutral gas phase of
a mixed population of galaxy types (e.g., Rao et al. 2003; Chen \&
Lanzetta 2003; Rao et al. 2011).

More recently, moderate resolution spectroscopy from the Sloan Digital
Sky Survey (SDSS; York et al.\ 2000) has led to a huge increase in the
number of quasar spectra available for QAL surveys.  This has advanced
the study of intervening absorption systems, moving sample sizes into
the realm of high statistical significance.  The survey of Nestor,
Turnshek, \& Rao (2005, hereafter NTR05) utilized $\sim 3400$ quasar
spectra from the SDSS early data release (EDR). This present contribution
is a continuation of the work begun in NTR05, and the reader is referred to 
NTR05 for additional details on the methods used here.
From the spectra analyzed by NTR05, a
statistical sample of over 1300 \ion{Mg}{2} systems having \Wb$\ge
0.3$\AA\ was derived.  In addition to providing a firm basis for the
statistics of \ion{Mg}{2}\ absorbers, that sample has also been used
for follow-up studies, such as the first identification of a
correlation between gas-phase metals and absorbing gas velocity spread
(Nestor et al.\ 2003), an investigation of the mean spatial extent and
photometric properties of \ion{Mg}{2} absorbing galaxies (Zibetti et
al. 2005), and an expanded investigation of DLAs at $z<1.65$ (Rao et
al. 2006). The \ion{Mg}{2} EDR statistics have also been compared with
\ion{Mg}{2} statistics derived from a radio-selected sample of quasars
in order to place an upper limit on the number of \ion{Mg}{2}
absorbers missed in optical surveys due to the dimming of background
quasars by dust in the intervening absorbers (Ellison et al. 2004;
Ellison \& Lopez 2009).  In the past few years, much larger \ion{Mg}{2} absorber
catalogs have been developed from subsequent data releases of the SDSS
(e.g., Prochter, Prochaska, \& Burles 2006; this 
contribution).  Our present catalog has already been used to study the
statistical properties of the absorbing gas and associated galaxies in
order to take advantage of the high precision inherent in such large
surveys. These studies include an expanded investigation of  the
correlation between metals and absorbing gas velocity spread
(Turnshek et al. 2005), an expanded investigation of the mean
spatial extent and photometric properties of the absorbing galaxies
(Zibetti et al. 2007), investigations of the sightline velocity
clustering of the absorbers and the transverse spatial clustering of
bright galaxies along the sightlines passing through them (Rimoldini
2007), an investigation of the dust extinction and gravitational
lensing effects that the absorbers induce on background quasars
(M\'{e}nard et al. 2008), and a sensitive search for emission
associated with the absorbing galaxies (M\'{e}nard et al. 2011).  In
addition, follow-up imaging to search for galaxies associated with the
rarest and strongest \ion{Mg}{2} systems has taken place (Nestor et
al.\ 2007, 2010).

In this paper we present the current state of our SDSS \ion{Mg}{2}
absorber catalog, which is presently complete through SDSS Data
Release Four (DR4).  As noted above, this contribution represents a continuation of
the work presented by NTR05 in that we used the method of NTR05; however, we
did not use their measurements of the EDR \ion{Mg}{2}
absorbers.\footnote{The current version of the catalog does not
include all EDR quasars.  It only includes those quasars which meet
our catalog's selection criteria (see \S2).} It contains $>17,000$
absorbers in the redshift interval $0.36 \lesssim z \lesssim 2.28$
identified in $\sim 44,600$ SDSS quasar spectra.  The catalog, along
with other information (\S4), is now publicly available at
http://enki.phyast.pitt.edu/PittSDSSMgIIcat.php. Searches of more
recent SDSS quasar spectra are underway, and in the future we hope to
extend the catalog to include additional \ion{Mg}{2} absorbers beyond
DR4. However, future extensions will only be announced and made
available on the Web site.  The combination of size and well understood
statistics makes the catalog ideal for (1) statistical studies of,
for example, the properties and evolution of the low-ionization and
neutral gas regions of galaxies, (2) identifying rare systems, which
are readily identified only in very large surveys, and (3) selecting
well-defined samples for follow-up studies. In a subsequent paper we
will present and discuss the statistical properties of the absorbers
in the present catalog (D. B. Nestor et al., in preparation).
 
\section{Quasar Selection Criteria}
\label{section:criteria}
The quasar spectra used to build the \ion{Mg}{2} absorber catalog are
a subset of the SDSS data releases through DR4. The
 searches for \ion{Mg}{2} absorption doublets 
were generally performed on the initially released versions of the spectra. However,
spectra have been re-extracted and re-calibrated in subsequent SDSS data releases 
(Adelman-McCarthy et al. 2008; Abazajian et al 2009). Therefore, for example, a quasar spectrum taken 
from DR7 may not be identical to the one we analyzed to form the catalog. However, since the nature of
measuring an absorption-line equivalent width involves fitting a continuum and making a measurement 
over a small wavelength interval, the re-extraction and re-calibration of spectra will generally have no 
effect on the statistical properties of the absorbers in the catalog. 

The
quasar spectra to be searched for \ion{Mg}{2} absorbers were isolated using an SQL
query of the public SDSS DR4 database (available at
http://cas.sdss.org/astrodr4/).  All of the spectroscopic and
photometric data in the SDSS are contained in two master tables: {\tt
SpecObjAll}, for spectroscopic data, and {\tt PhotoObjAll}, for
photometric data.  The data in these two tables are organized into
numerous ``views'' by imposing an array of selection criteria on the
master tables.  We used a  {\tt SpecPhoto} view to generate the
quasar sample to be searched.  The view we used was intended to eliminate all spectra of
poor quality, sky spectra, duplicate observations (those spectra
labeled as SECONDARY), and all data outside the primary DR4 survey
area. In addition, {\tt SpecPhoto} contains all of the spectroscopic
information and the most accurate photometric information at the time of the data release (those
observations labeled as BEST) for each included object.

We used four criteria to define the quasar sample: (1) a spectral
classification of 3 or 4, which corresponds to {\tt specClass = QSO} or {\tt HIZ\_QSO},
respectively, (2) $z > 0.36 $ to ensure visibility of the \ion{Mg}{2}
$\lambda\lambda$2796,2803 doublet at the SDSS low wavelength limit of
3820 \AA, (3) $zstatus > 1$ to eliminate missing or failed redshift
measurements, and (4) $i< 20$ for the fiber magnitude.  Incorporating
these criteria into an SQL query of the {\tt SpecPhoto} view 
yielded $\sim 44,600$ quasars.  These quasars form the basis for the 
\ion{Mg}{2} absorber catalog.  The quasars that meet our redshift, magnitude,
and zstatus selection criteria represent $\sim 72\%$ of all objects in
SDSS DR4 which are spectroscopically classified as {\tt QSO} or {\tt HIZ\_QSO}.
Alternatively stated, $\sim 28\%$ of SDSS DR4 quasars were unsuitable
for inclusion in the catalog because of emission redshifts that were
too low or poorly constrained, or magnitudes that were too faint for
recording spectra of sufficient quality.

\section{Construction of the Catalog}
\label{section:construction}
There were three main steps which led to the construction of the
\ion{Mg}{2} catalog: (1) automated processing of the quasar spectra,
(2) visual inspection of the automatically identified \ion{Mg}{2}
candidate doublets to eliminate systems judged to be unreliable, and
(3) measurement of the confirmed \ion{Mg}{2} doublets.  NTR05
developed this procedure and applied it to the SDSS EDR data.  In this
section, we briefly review these steps, referring the reader to NTR05
for more detail. When we generically refer to an absorption line's wavelength ($\lambda_0$), 
rest equivalent width ($W_0$), or error in rest equivalent width ($\sigma_{W_0}$), 
this applies to the particular line in question (e.g., see Table 1). 

\subsection{Automated Processing of Quasar Spectra}
The automated processing of the quasar spectra is comprised of two
distinct phases: the normalization of the spectra and identification of
\ion{Mg}{2} doublet candidates.

To maximize computation and storage efficiency, the data were abridged
to exclude spectral regions not in the range to be searched for
\ion{Mg}{2} systems. This included the spectral regions shortward of
the quasars' Ly$\alpha$ emission lines, in order to avoid the
confusion and inaccuracy associated with searching in the Ly$\alpha$
forest, and longward of \ion{Mg}{2} $\lambda$2800 broad emission lines,
where intervening \ion{Mg}{2} absorption would be absent.  A
combination of cubic splines and Gaussians was employed to fit the
spectra, including the true continua and broad emission and broad
absorption features.  These continua fits, or pseudocontinua fits 
(i.e., continua that fit broad emission or absorption features and 
not just true continua), were
used to normalize the spectral fluxes and flux uncertainties.  The
continuum-fitting process was successful for the vast majority 
($\gg 99\%$) of spectra.  The software only failed to derive satisfactory continua for
a small number of quasars having rare spectra, such as extreme broad
absorption-line (BAL) quasars; these spectra were excluded from the
sample.  Occasionally the automated software produced a poor fit in a
localized spectral region.  Such occurrences were very infrequent in
comparison to the overall size of the survey, and always produced
false candidates (rather than missing true systems) when
present. These cases had no measurable effect on the overall
properties of the catalog.  In the uncommon circumstance (i.e., in $< 1\%$ of the cases) of a 
poor continuum fit being coincident with an \ion{Mg}{2} absorption doublet, the fit was
interactively re-fit to produce a satisfactory continuum before final equivalent 
width measurements were performed (see \S3.3).

All normalized spectra were run though a routine that flags possible
\ion{Mg}{2} doublets.  In order to isolate a sample of absorbers that were primarily 
intervening, systems with measured absorption redshifts
within 3000 km s$^{-1}$ of the SDSS quasar emission redshift were not
included in the catalog. Therefore, biases that might arise due to
overdensities of absorbers in quasar environments will be minimized
or avoided in any statistical study based on the catalog.  This same
software also determines the $\lambda2796$ rest equivalent width detection limit, \Wl\
(see \S\ref{section:meas}), as a function of redshift or observed wavelength, 
$(1+z)=\lambda_{obs}/2796.352$, for the range of \ion{Mg}{2} redshifts 
corresponding to a searched spectral region.  This
\Wl\ information was recorded for each wavelength in each searched spectral 
region so that it could be integrated 
over all sightlines to determine the total number of redshift intervals 
surveyed as a function of \Wl\ and
redshift (e.g., see Figure 8 in NTR05).  

The \ion{Mg}{2} doublet finding routine is conservative by
design to ensure that all possible doublets are located in this initial step.   A total of
$\sim 44,600$ quasar spectra were searched for the \ion{Mg}{2}
absorption doublet.  The conservative nature of our automated doublet
finder led to a large number of flagged candidates: on average, more
than two \ion{Mg}{2} candidates were flagged for each quasar.

\subsection{Visual Inspection of Candidate \ion{Mg}{2} Doublets}
\label{section:inspection}

All candidate \ion{Mg}{2} doublets were individually inspected by eye
to determine their validity.  As the candidate-flagging routine was
designed to be conservative, this resulted in the rejection of
approximately 85\% of all candidates during visual
inspection.\footnote{The high rate of visual rejection may have
unfortunately resulted in the accidental rejection of some clearly
significant candidates (see \S4.2).}  The most common false positives
were due to: (1) absorption in a BAL trough, (2) multiple
identifications of a candidate \ion{Mg}{2} doublet (this sometimes
occurred within strong \ion{Mg}{2} absorption profiles; 
we selected the apparent best absorption redshift and
rejected redundant identifications), (3) non-\ion{Mg}{2} absorption
lines, or (4) coincidentally aligned noise spikes.

Consistent with previous QAL surveys (\S1), our experience with the SDSS spectra 
has shown that isolated
absorption features measured with a significance $\ge 5\sigma_{W_T}$ are
what the eye determines to be confident detections.  
Here $\sigma_{W_T}$ is the theoretical error in rest equivalent width used to calculate the
detectability limit, \Wl,  of an \ion{Mg}{2} doublet (see \S \ref{section:meas}).
When combined with additional information, such as the confident detection
of a doublet partner, significances $\ge 3\sigma_{W_T}$ tend to
be consistent with confident-by-eye detections.  Candidate doublets
identified with significances close to these thresholds would often be
difficult to categorize as real or spurious based on quick visual
inspection.  Such candidates were generally retained at this stage by
default.  However, after final measurements were available (see \S3.3),
we used visual inspection to reject suspicious candidates. This included
candidates with doublet ratios significantly different from the
physically-allowed range of 1.0 $\le$ \Wb/\Wr $\le$ 2.0. We also examined 
each spectrum to assess the possibility that a candidate \ion{Mg}{2} system
was flagged due to the presence of a QAL system at
another redshift, and candidates which were not convincingly \ion{Mg}{2}
were rejected.

\subsection{Measurements of \ion{Mg}{2} Doublets}
\label{section:meas}

The reported measurements of the redshifts and rest equivalent
widths of all accepted systems were finalized interactively when necessary.  A pair of
Gaussian line profiles separated in wavelength by the redshifted doublet separation
was fitted to each accepted candidate's profile.  The location of
these fitted profiles defined the observed redshift and was used to
perform an optimal extraction (see NTR05) of each line's rest
equivalent width, $W_0$, and its associated error, $\sigma_{W_0}$.  This is the optimal
procedure when the observed profile is well described by the chosen
function. We draw particular attention to Equations (1) and (2) from NTR05:
\begin{equation}
(1+z)\,W_0 = \frac{\sum_i P(\lambda_i-\lambda_0)\, (1-f_i)}{\sum_i
P^2(\lambda_i-\lambda_0)}\,\Delta\lambda,
\label{measureW}
\end{equation} 
\begin{equation}
(1+z)\,\sigma_{W_0} = \frac{\sqrt{\sum_i P^2(\lambda_i-\lambda_0)\,
\sigma_{f_i}^2}}{\sum_i P^2(\lambda_i-\lambda_0)}\,\Delta\lambda,
\label{measuresigma}
\end{equation} 
where $P(\lambda_i-\lambda_0)$, $\lambda_i$, $f_i$, and $\sigma_{f_i}$ 
represent the line profile centered at $\lambda_0$, the wavelength, 
the normalized flux, and flux uncertainty as a function of pixel.  The sum is 
performed over an integer number of pixels that cover at least $\pm 3$ 
characteristic Gaussian widths.  For systems with clearly saturated or resolved profiles,
two pairs of Gaussians were used to fit the absorption system, and
this proved to be almost always successful at accurately reproducing
the observed profiles.  In these cases the reported redshift 
is the rest-equivalent-width-weighted redshift.
For weaker and/or unresolved systems, the
additional degrees of freedom afforded by a double-Gaussian did not
generally give different values for \Wb\ and \Wr.  In over 95\% of the
cases, doublet members were fit with single-Gaussian profiles.  Upon careful
inspection of the systems retained in the catalog after measurement,
it was found that systems stronger than \Wb\ $\gtrsim 3$ \AA\
frequently had their strengths slightly ($< 1\sigma_{W_0}$) but
systematically overestimated when using a single-Gaussian profile
fit. Therefore,  the $\sim 700$ strongest systems were re-inspected
and re-measured using double Gaussians when appropriate. The 
centroids of each Gaussian were left free during the fit.

Consistent with experience (\S\ref{section:inspection}), the values for the 
$\lambda2796$ rest equivalent width detection limit, \Wl, as a function of 
redshift or observed wavelength, $(1+z)=\lambda_{obs}/2796.352$, were calculated  
by requiring \Wb\ $\ge 5\sigma_{W_T}$ and \Wr\ $\ge 3\sigma_{W_T}$.
Thus, to compute \Wl\ for each spectrum we did the following. 
We started by determining the $1\sigma$ rest
equivalent width error, $\sigma_{W_0}$ (Equation (2)), for a theoretical,
unresolved line near the location of each doublet member and set $\sigma_{W_T} = \sigma_{W_0}$.  
We then took the value of \Wl\  to be the larger of five times this error
($5\sigma_{W_T}$) at the location of the $\lambda2796$ line or six times
this error ($2\times 3\sigma_{W_T}$) at the location of the
$\lambda2803$ line.  The factor of two avoids introducing a detection
bias caused by the \ion{Mg}{2} doublet ratio, which has a theoretical
range of approximately 1.0 (completely saturated) to 2.0
(completely unsaturated).  Systems with measured \Wb $<$ \Wl\ were excluded from
the catalog.  This eliminated the vast majority of suspicious
systems that were not rejected during visual inspection.  At this
point, $\sim 17,000$ \ion{Mg}{2} doublets were considered
real and successfully measured.  Whenever possible, systems close in
redshift were measured as separate systems.  It is important to note
that the redshift difference for which this was possible was not
constant, because it depended on both redshift and the widths of the features,
and therefore on \Wb\ and \Wr.  We retained such systems as separate
in the catalog and note that they can always be combined {\it ex post
facto} for statistical purposes, when necessary.  This is a deviation
from the policy of NTR05, where systems with separations less than 500
km s$^{-1}$ were always measured as a single system.

\subsection{Measurements of Adjacent \\ \ion{Mn}{2}, \ion{Fe}{2}, and 
\ion{Mg}{1} Absorption Lines in Cataloged \ion{Mg}{2} Systems}
\label{section:otherabs}

In addition to the measurements of
\ion{Mg}{2}  $\lambda\lambda$2796,2803 doublets, we also made rest
equivalent width measurements at the predicted locations of six
additional absorption lines 
(\ion{Fe}{2} $\lambda\lambda 2586,2600$,
\ion{Mn}{2} $\lambda\lambda 2576,2594,2606$,
 and
\ion{Mg}{1} $\lambda2852$) for all of the systems in the
catalog, when possible. Since these lines are not too far displaced from the
\ion{Mg}{2} doublet, it was usually possible to make measurements at
their predicted locations. There are many significant detections. For example, for \ion{Mg}{2} systems 
in the redshift interval $0.36 \le z \le 2.28$ in SDSS spectra, the \ion{Mn}{2} $\lambda 2576$ transition 
is usually covered when $z>0.48$ and the \ion{Mg}{1} $\lambda 2852$ transition is
usually covered when $z<2.22$.  Including non-detections caused by lack of coverage,
the  $2\sigma$ and $3\sigma$ detection fractions for these six absorption transitions are as follows:
\ion{Fe}{2} $\lambda2586$ (53.7\% and 43.6\%), \ion{Fe}{2} $\lambda2600$ (73.5\% and 64.1\%), 
\ion{Mn}{2} $\lambda2576$ (10.9\% and 4.9\%),  
\ion{Mn}{2} $\lambda2594$ (8.1\% and 3.3\%), 
\ion{Mn}{2} $\lambda2606$ (7.0\% and 2.7\%), 
and \ion{Mg}{1} $\lambda2852$ (40.6\% and 27.9\%).

One motivation for making these additional measurements is to
facilitate future use of this catalog to identify candidate DLA systems.  
For example, as discussed by Rao et al. (2006), it appears that 
measurements of \ion{Fe}{2} $\lambda2600$, \ion{Mg}{2} $\lambda2796$, 
and \ion{Mg}{1} $\lambda2852$ can be used together to isolate a subset of
\ion{Mg}{2} systems that will contain a complete sample of DLA
systems, but with a higher DLA fraction than previously achieved. For example, 
follow-up UV spectroscopy to measure $N_{HI}$ in a properly
selected subset of \ion{Mg}{2} systems (see Rao et al. 2006) indicates
that the sample will be complete and $\sim42$\% will be DLAs.

\begin{figure}[h!]
\centerline{
\includegraphics[width=0.8\columnwidth,clip,angle=270]{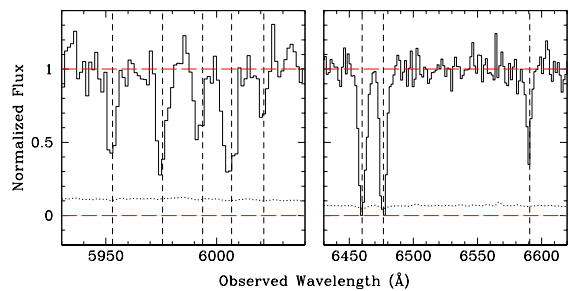}}
\caption{Two regions of the normalized spectrum of SDSS quasar
J085244.74$+$343540.4 showing eight metal absorption-line transitions
at $z=1.3102$ (\ion{Mn}{2} $\lambda2576$, \ion{Fe}{2} $\lambda2586$,
\ion{Mn}{2} $\lambda2594$, \ion{Fe}{2} $\lambda2600$,
\ion{Mn}{2} $\lambda2606$, \ion{Mg}{2} $\lambda\lambda2796$,2803, and
\ion{Mg}{1} $\lambda2852$). The lower dotted curve is the 1$\sigma$
error in normalized flux. }
\label{fig:spectrum}
\end{figure}

Figure \ref{fig:spectrum} shows two regions of a spectrum containing
one of the \ion{Mg}{2} absorption-line systems in the catalog. The
absorption system is in the spectrum of quasar SDSS
J085244.74$+$343540.4 ($z_{em}=1.655$) and the absorption redshift is
$z=1.310$. The figure shows the \ion{Mg}{2} doublet and the six
additional metal-line absorption transitions that have been measured
for this system. Table  \ref{table:metal-lines} presents
the metal-line measurements for this system as an example of
information that may be found in the catalog.

\section{The Catalog Characteristics}
\label{section:format}
\subsection{The Catalog Format}
The Pittsburgh SDSS \ion{Mg}{2} Quasar Absorption-Line Survey Catalog
is available for public use at
http://enki.phyast.pitt.edu/PittSDSSMgIIcat.php. It includes
information about the quasars that were searched for \ion{Mg}{2}
doublets, the measurements of the \ion{Mg}{2} doublets (\S3.3), 
the measurements of up to six additional metal absorption lines (\S3.4), and the
quasar spectral files in text format.  Values of \Wl\ (\S3.3) have 
been merged with the quasar spectral files, so the available
quasar spectral files contain flux, flux uncertainty, pseudo-continuum
fit, and \Wl, all as a function of wavelength.  The quasar properties in
Table \ref{table:format} are available for each quasar that was
searched for \ion{Mg}{2}.  Table \ref{table:format} is drawn from DR4,
and we refer the reader to http://www.sdss.org/dr4 for further
explanation of the values given in the table.

In summary, quasar spectra files and DR4 quasar properties are made available
for $\sim 44,600$ quasars, and absorption system information is
available for $> 17,000$ \ion{Mg}{2} doublets.

\subsection{Missed Systems}
We have attempted to quantify our incidence of missed \ion{Mg}{2}
absorption doublets which are real, clearly significant, and located
within our searched quasar sample.  There are two possible steps
during the construction of the catalog where one might think that an \ion{Mg}{2} system
could be missed or lost: (1) the automated \ion{Mg}{2} identification
routine and (2) the interactive inspection and/or measurement of a
candidate \ion{Mg}{2} system which often results in removing a system
for the reasons given in \S3.2 and \S3.3. As the \ion{Mg}{2}
identification software has been robustly tested and conservatively
written (see NTR05), we are confident that the first possibility is an
exceedingly small source of missed \ion{Mg}{2} absorbers when compared
to the second possibility, which includes the potential for human
error.

To quantify our incidence of missed \ion{Mg}{2} absorption systems, we
randomly selected 500 quasars from our catalog.  Two of us (A.M.Q and
D.B.N) then visually inspected 200 separate spectra and 100 common
spectra (for a total of 500 spectra) to search for \ion{Mg}{2}
absorbers.  As explained in \S\ref{section:inspection}, our formal
detection/rejection threshold is such that the human eye seems to be
generally capable of identifying \ion{Mg}{2} absorbers to slightly
better than the sensitivity limit (\Wl) adopted for the construction
of the catalog.  Therefore, manually re-inspecting the spectra for
\ion{Mg}{2} absorption doublets in a slow and careful manner allows us
to account not only for software errors, but also human errors made
during the initial laborious process of manually accepting/rejecting
candidates from the extremely large master list.  In the entire sample
of 500 spectra, we discovered five significant missed systems with \Wb
$-$ \Wl $> \sigma_{W_0}$. Therefore, we conclude that human error caused
the rejection of some systems. We estimate that for every 100 searched
spectra approximately one \ion{Mg}{2} absorber unambiguously above our
detection threshold has been missed. This is a systematic human error that 
appears not to be correlated with any \ion{Mg}{2} system property.

\subsection{Quasar Properties}

\begin{figure}[h]
\vspace{0.25cm}\centerline{
\includegraphics[width=0.75\columnwidth,clip,angle=0]{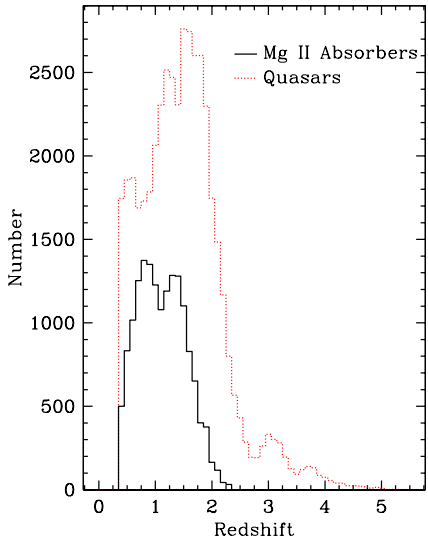}}
\caption{Emission redshift distribution for all quasars in the
catalog (red dotted line) and the absorption redshift distribution for
all \ion{Mg}{2} systems in the catalog (black solid line).  The
redshift bin width for both distributions is 0.1.  The emission
redshift range of the quasars in the catalog is constrained by the
criteria given in \S2. The absorption redshift range of the
\ion{Mg}{2} systems is determined by the wavelength coverage of SDSS
spectra, namely $0.36 \lesssim z \lesssim 2.28$. The drop in the
number of absorption systems near $z\approx1.1$ is due to a decrease
in SDSS spectrograph sensitivity in the transition region from the
blue-side to red-side cameras.}
\label{fig:zdist}
\end{figure}

\begin{figure}[h]
\centerline{
\includegraphics[width=0.85\columnwidth,angle=0]{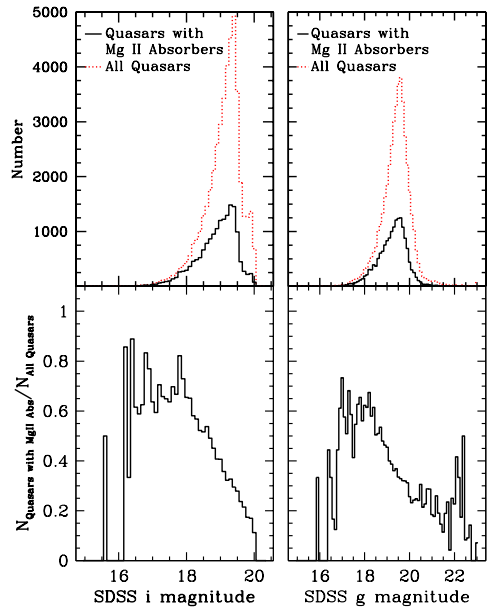}}
\caption{Top panels show the $i$ (left) and $g$ (right) 
magnitude distributions for all quasars in the catalog (red dotted line) and for those 
quasars with one or more \ion{Mg}{2} absorbers (black solid line).  The magnitude bin width is 0.1.  
The bottom panels show the 
corresponding fractions of quasars with one or more \ion{Mg}{2} systems as a function of 
magnitude, without regard to quasar redshift. As expected, the brighter quasars have a 
higher probability of exhibiting \ion{Mg}{2} absorption since the high S/N of their spectra
permits the detection of weaker \ion{Mg}{2} systems, which have a higher incidence than stronger systems.}
\label{fig:QSOmag}
\end{figure}

Figure \ref{fig:zdist} shows the distribution of SDSS emission
redshifts for all quasars in the catalog along with the distribution
of absorption redshifts for all \ion{Mg}{2} systems in the catalog.
The distribution of emission redshifts is similar to the one derived
from the DR5 quasar catalog (Schneider et al. 2007).\footnote{A DR4
quasar catalog was not published.} Schneider et al. (2007) note that
the dips in the emission redshift distribution near $z = 2.7$ and $z =
3.5$ are due to the SDSS quasar selection algorithm.  Figure
\ref{fig:QSOmag} shows the SDSS $i$ magnitude (left panel) and $g$
magnitude (right panel) distributions for all quasars in the catalog
(red dotted line) and for those quasars with at least one intervening
\ion{Mg}{2} absorption system (black solid line).  The $i$ magnitude
distribution is similar to the one published by Schneider et
al. (2007) for  DR5 quasars. The bottom panels of Figure \ref{fig:QSOmag}
show the corresponding fractions of quasars with one or more \ion{Mg}{2}
systems as a function of magnitude, without regard to quasar redshift. 
As expected, the brighter quasars have a higher probability of exhibiting
\ion{Mg}{2} absorption since the high signal-to-noise ratio of their 
spectra permits the detection of weaker \ion{Mg}{2} systems, which have
a higher incidence than stronger systems (e.g., NTR05). 

\subsection{\ion{Mg}{2} Absorber Properties}
A thorough treatment of the \ion{Mg}{2} absorber statistics from the
catalog will be presented in a  subsequent paper (D. B. Nestor et al., in
preparation).  Here, we present the raw properties of the \ion{Mg}{2}
doublets in the catalog to illustrate the properties of the catalog
itself.

Returning to Figure \ref{fig:zdist}, the black histogram shows the
distribution of absorber redshifts for \ion{Mg}{2} doublets in the
catalog.  The redshift distribution of the \ion{Mg}{2} absorbers is
constrained to lie between $0.36 \lesssim z \lesssim 2.28$ because of
the wavelength coverage of the SDSS spectrograph.  As mentioned in
NTR05, the dip at $z \approx 1.1$ is due to the ubiquitous decrease in
signal-to-noise ratio at the wavelength corresponding to the split
between the blue-side and red-side cameras by the dichroic.   The
black histograms in the top panels of Figure \ref{fig:QSOmag} show the distribution of
$i$ and $g$ magnitudes for quasars having at least one \ion{Mg}{2}
absorber in its spectrum. As noted in \S4.3, the bottom panels 
indicate that a higher fraction of \ion{Mg}{2} systems have been detected
in brighter quasars. This is due to the fact that the higher signal-to-noise
ratios of the brighter quasars permit the detection of weaker \ion{Mg}{2}
systems, which have a higher incidence than stronger \ion{Mg}{2} systems.

\begin{figure}[h]
\centerline{
\includegraphics[width=0.75\columnwidth,angle=0]{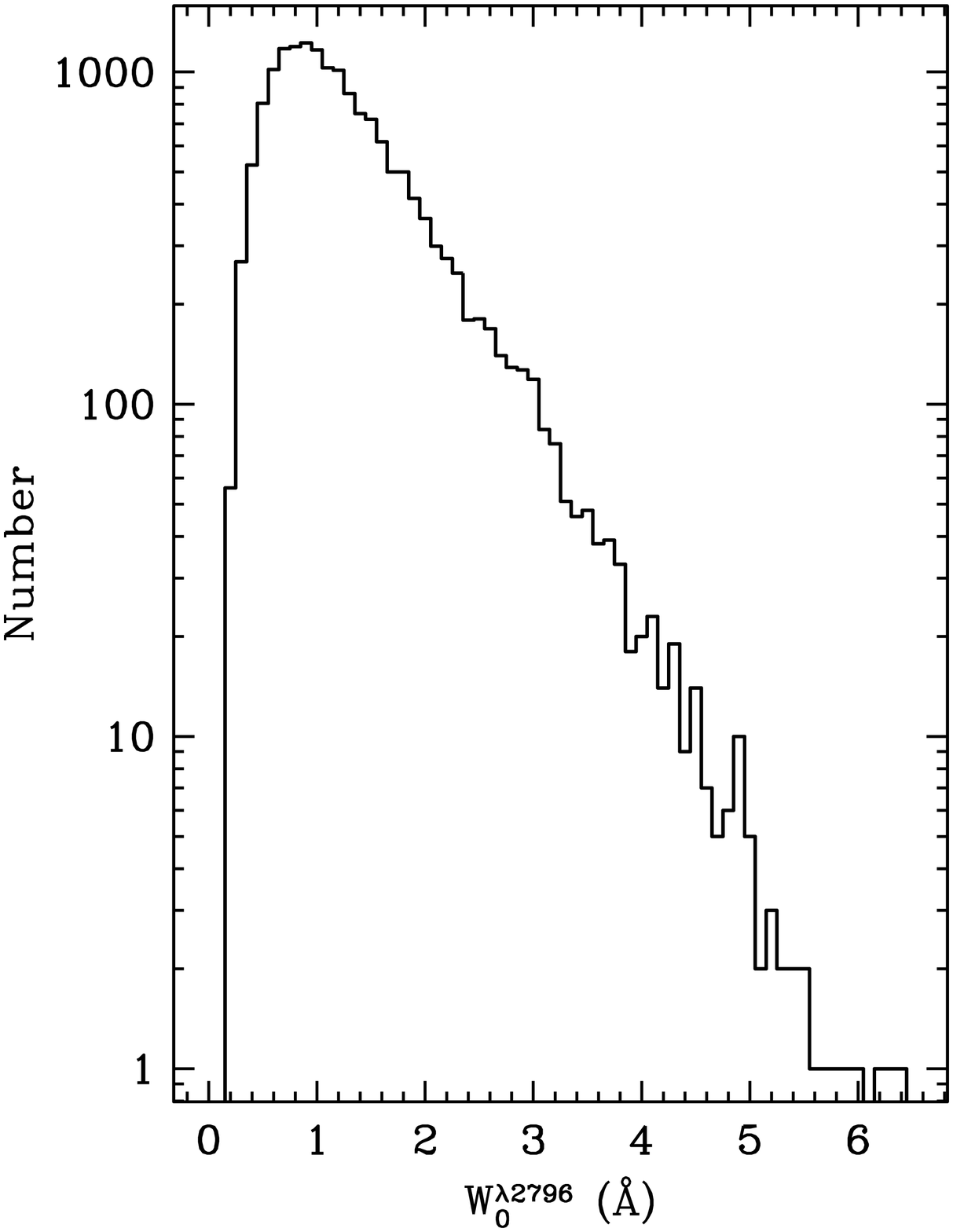}}
\caption{Observed $\lambda2796$ rest equivalent width, \Wb, distribution for the \ion{Mg}{2} 
absorbers in the catalog. The bin width is 0.1 \AA.}
\label{fig:REW2796dist}
\end{figure}

The observed \Wb\ distribution is presented in Figure \ref{fig:REW2796dist}.
Nearly 50\% of the absorbers have $0.6$ \AA\ $<$  \Wb\ $< 1.2$ \AA.  We
identify $\sim 10,000$ \ion{Mg}{2} absorbers with \Wb\ $>$ 1.0 \AA.  For
comparison, Prochter et al. (2006) compiled 7,421 \ion{Mg}{2}
absorbers with \Wb $>$ 1.0 \AA\ in their SDSS DR3 catalog. A large
fraction of the absorbers in our catalog have high rest equivalent
width, with $\sim$16\% having \Wb $>$ 2.0 \AA.

\begin{figure}[h]
\centerline{
\includegraphics[width=1.0\columnwidth,angle=270]{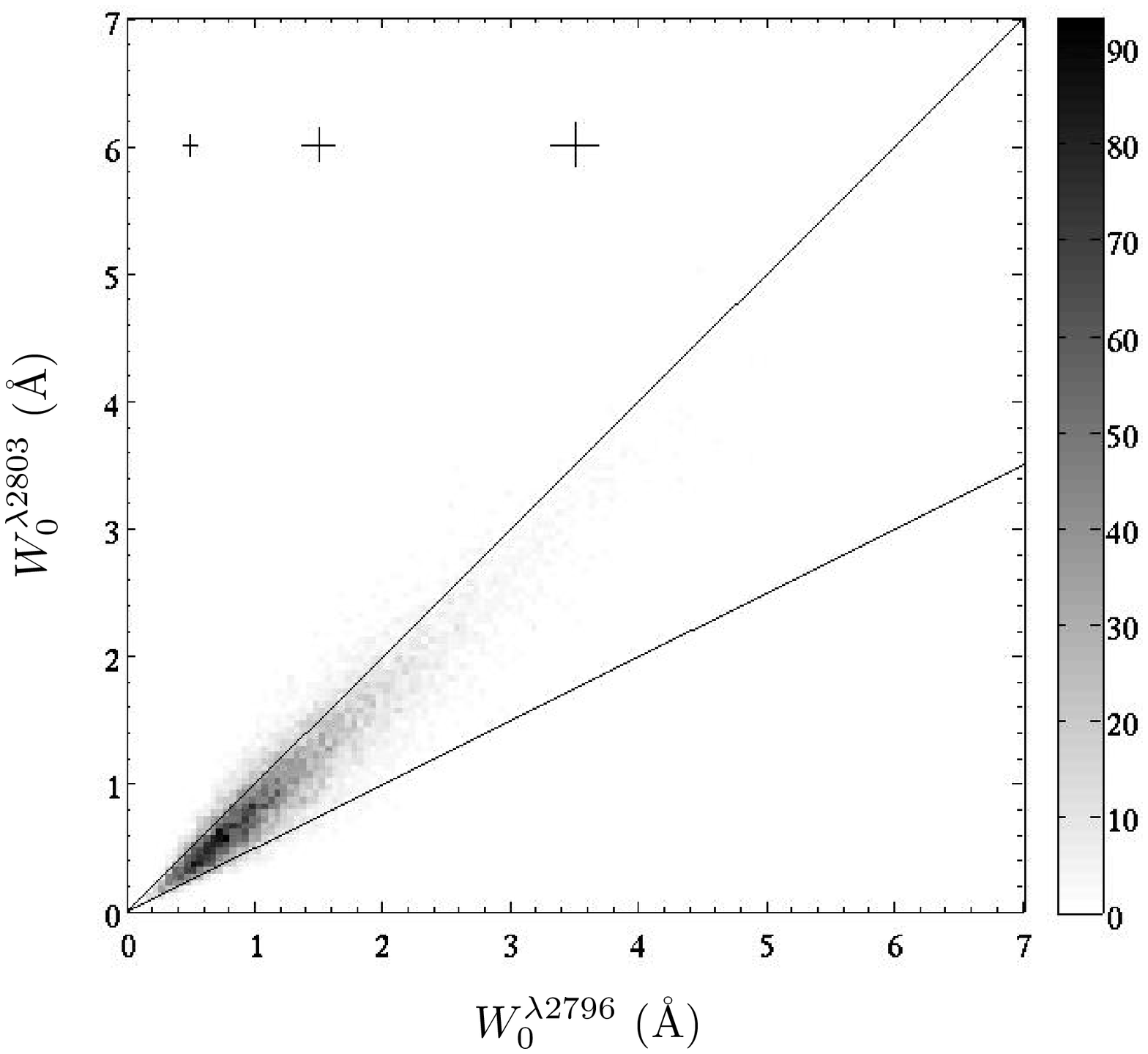}}
\caption{Plot of \Wb\ vs. \Wr\ for absorbers in the catalog is shown.  
The two diagonal lines represent the theoretical boundaries for the  
\ion{Mg}{2} doublet ratios in a hypothetical error-free data set (i.e., 
$1 \leq$ \Wb\ $\backslash$ \Wr\ $\leq 2$).  The vast majority of doublets 
in the catalog have doublet ratios that fall within the theoretical boundaries. 
Those that fall outside the theoretical boundaries can be attributed to 
errors associated with noise and/or line blending.  The crosses in the upper 
left corner represent the average {$\pm 1\sigma_{W_0}$} errors in the \Wb\ and 
\Wr\ measurements for \Wb\ = \Wr\ $\approx 0.5$ \AA, 1.5 \AA, and 3.5 \AA, 
respectively. The gray-scale bar on the right-hand side provides information 
on the frequency distribution of doublet ratios.}
\label{fig:REW2796_REW2803}
\end{figure}

\begin{figure}[h]
\centerline{
\includegraphics[width=0.75\columnwidth,angle=0]{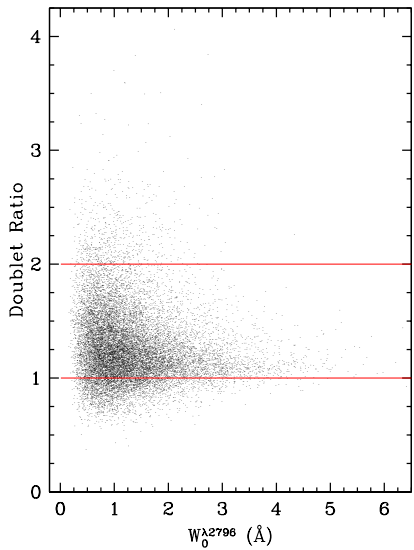}}
\caption{Plot of \Wb/ \Wr\ vs. \Wb\ for the \ion{Mg}{2} absorption doublets 
in the catalog. This is a different rendering of the same data shown in Figure 5.}
\label{fig:DR_REW2796}
\end{figure}

Figures \ref{fig:REW2796_REW2803} and \ref{fig:DR_REW2796} show the
measured relationship between \Wb\ and \Wr.  The diagonal lines in
Figure \ref{fig:REW2796_REW2803} represent the theoretical limits for
completely saturated (\Wb$/$\Wr$= 1.0$) and unsaturated
(\Wb$/$\Wr$= 2.0$) absorption.  The error bars along the
horizontal axis represent the average measured $\sigma_{W_0}$ for
various rest equivalent width regions.  Figure \ref{fig:DR_REW2796}
illustrates the relationship between the doublet ratio and the value
of \Wb.

The vast majority of the \ion{Mg}{2} absorbers in the catalog have
doublet ratios that lie within the theoretical limits, and deviations
from these limits can be attributed to noise and line blending.  In
some instances, for example, systems that are clearly real, exhibiting
high-significance metal lines from other species at the same
absorption redshift, will have poorly measured \Wr\ values due to
contamination from a line in another absorption system. Such systems
may have measured doublet ratios well outside of the theoretical
range.  Most of our effort has been given to obtaining the best
possible measurements of \Wb\ in identified systems. Therefore, these
values can generally be used with more confidence.  Of course, we also
attempt to measure an accurate value for \Wr, but less effort and no
testing has been devoted to determining possible problems with these
measurements. Therefore, more caution should be exercised if the
desire is to use the \Wr\ values in high-precision studies.

\section{Summary}
We present the Pittsburgh SDSS \ion{Mg}{2} Quasar Absorption-Line
Survey Catalog $-$ a well-characterized catalog of \ion{Mg}{2}
absorption doublets found in the spectra of quasars up through SDSS
DR4, along with supporting information.  We have combined automated
and manual techniques to search the spectra of $\sim 44,600$ quasars
to arrive at a catalog of $>17,000$
\ion{Mg}{2} $\lambda\lambda$2796,2803 absorption doublets which we now
make available for public use. The available information includes the
\ion{Mg}{2} absorber catalog itself, with measurements of the
absorbers and associated errors, measurements of up to six nearby
metal absorption lines (\ion{Mn}{2}, \ion{Fe}{2}, and \ion{Mg}{1}),
and information on the quasar catalog which was searched, including
wavelength-dependent spectral files (flux, flux uncertainty, pseudo-continuum fit, and  
$\lambda2796$ rest equivalent width detection limit) for each quasar. The
results contain the information necessary for statistical studies.  We
estimate that for every 100 searched  spectra approximately one
\ion{Mg}{2} absorber unambiguously above our  detection threshold has
been missed.  A complete analysis of the statistics of the catalog
will be given in a subsequent paper (D. B. Nestor et al., in preparation).

We make the catalog available to the community with the hope that its
very large size and documented statistical characteristics will lead
to future innovative investigations at low to moderate redshift.  Any
future extensions to the catalog, which would include data beyond SDSS
DR4, will only be announced and
made available at the catalog's Web site
http://enki.phyast.pitt.edu/PittSDSSMgIIcat.php.  Please email queries
regarding the catalog to mgiicat@pitt.edu.

\acknowledgments

AMQ is supported by a Marshall Scholarship and a National Science
Foundation Graduate Research Fellowship. DAT and SMR acknowledge
support from NSF grant AST-0307743.  AMQ and ANW acknowledge REU
support from NSF grants AST-0307743 and PHY-0244105.  DBN acknowledges
early support from Zaccheus Daniel and Andrew Mellon pre-doctoral
Fellowships at the University of Pittsburgh and support from the
STFC-funded Galaxy Formation and Evolution programme at the Cambridge
Institute of Astronomy.

We thank Eric Furst and Daniel Owen for their assistance with the
visual inspection of candidate \ion{Mg}{2} doublets. The involvement
of these undergraduate students was made possible by REU support from
NSF grants AST-0307743 and PHY-0244105.
 
We are grateful to the anonymous referee for helpful comments.
We thank members of the SDSS collaboration who made the project a
success. 

Funding for the creation and distribution of the SDSS Archive has
been provided by the Alfred P. Sloan Foundation, Participating
Institutions, NASA, NSF, DOE, the Japanese Monbukagakusho, and the
Max-Planck Society. The SDSS Web site is http://www.sdss.org. The SDSS
is managed by the Astrophysical Research Consortium for the
Participating Institutions: the American Museum of Natural History,
Astrophysical Institute Potsdam, University of Basel, Cambridge
University, Case Western Reserve University, University of Chicago,
Drexel University, Fermilab, the Institute for Advanced Study, the
Japan Participation Group, Johns Hopkins University, the Joint
Institute for Nuclear Astrophysics, the Kavli Institute for Particle
Astrophysics and Cosmology, the Korean Scientist Group, the Chinese
Academy of Sciences (LAMOST), Los Alamos National Laboratory, the
Max-Planck-Institute for Astronomy (MPIA), the Max-Planck-Institute
for Astrophysics (MPA), New Mexico State University, Ohio State
University, University of Pittsburgh, University of Portsmouth,
Princeton University, the United States Naval Observatory, and 
University of Washington

\begin{table}
\centering 
\caption{\textsc{Metal-Line Measurements for J085244.74$+$343540.4
($z_{em}=1.655$) }}
\begin{tabular}{lcc}
\\ \hline \hline Transition & $\lambda_0$ & $W_0^1 \pm {\sigma_{W_0} }$ \\ &
(\AA)                & (\AA)        \\ \hline 
\ion{Mg}{2}   &  2796.35   & $3.046^2\pm0.083$ \\ 
\ion{Mg}{2}   &  2803.53   & $2.930\pm0.085$ \\ 
\ion{Fe}{2}    &  2586.65   & $1.567\pm0.159$ \\
\ion{Fe}{2}    & 2600.17   &  $2.189\pm0.166$ \\ 
\ion{Mn}{2}   &  2576.88 &  $1.261\pm0.138$ \\ 
\ion{Mn}{2}   &  2594.50   &  $1.034\pm0.179$ \\ 
\ion{Mn}{2} &  2606.46   & $0.583\pm0.165$    \\ 
\ion{Mg}{1}   & 2852.96   &  $1.150\pm0.087$   \\  
\hline $^1 z_{abs} = 1.3102$\\
$^2$ \Wl\ = 0.372 \AA
\label{table:metal-lines}
\end{tabular}
\end{table}

\begin{table}
\centering 
\caption{\textsc{SDSS DR4 Quasars Searched for \ion{Mg}{2}}}
\begin{tabular}{cl}
\\ \hline \hline Column & Description\\ \hline 
1 & J2000 name\\ 
2 & SDSS Modified Julian Date (MJD)\\
3 & SDSS Plate Number \\
4 & SDSS Fiber Number \\ 
5 & Right ascension (J2000)\\ 
6 & Declination (J2000)\\ 
7 & z$_{em}$\\ 
8 & Error in z$_{em}$\\ 
9 & BEST fiber $u$ magnitude\\ 
10& BEST fiber $g$ magnitude\\ 
11& BEST fiber $r$ magnitude\\ 
12& BEST fiber $i$ magnitude\\ 
13& BEST fiber $z$ magnitude\\ 
14& Error in BEST fiber $u$ magnitude\\ 
15 & Error in BEST fiber $g$ magnitude\\ 
16 & Error in BEST fiber $r$ magnitude\\ 
17 & Error in BEST fiber $i$ magnitude\\ 
18 & Error in BEST fiber $z$ magnitude\\
\hline
\label{table:format}
\end{tabular}
\end{table}

\end{document}